\documentclass[aps,prd,twocolumn,superscriptaddress,showpacs,floatfix,nofootinbib]{revtex4-1}

\usepackage[dvipdfmx]{graphicx}
\usepackage{hyperref}
\usepackage{color}
\usepackage{amsmath}
\usepackage{comment}
\hypersetup{
	colorlinks=true, 
	pdfborder={0 0 0},
	linkcolor=blue,
	citecolor=blue,
	urlcolor=blue,
}

\begin{document}

\title{Search for proton decay into three charged leptons in 0.37 megaton-years exposure of the Super-Kamiokande}

\newcommand{\AFFicrr}{\affiliation{Kamioka Observatory, Institute for Cosmic Ray Research, University of Tokyo, Kamioka, Gifu 506-1205, Japan}}
\newcommand{\AFFkashiwa}{\affiliation{Research Center for Cosmic Neutrinos, Institute for Cosmic Ray Research, University of Tokyo, Kashiwa, Chiba 277-8582, Japan}}
\newcommand{\AFFipmu}{\affiliation{Kavli Institute for the Physics and
Mathematics of the Universe (WPI), The University of Tokyo Institutes for Advanced Study,
University of Tokyo, Kashiwa, Chiba 277-8583, Japan }}
\newcommand{\AFFmad}{\affiliation{Department of Theoretical Physics, University Autonoma Madrid, 28049 Madrid, Spain}}
\newcommand{\AFFubc}{\affiliation{Department of Physics and Astronomy, University of British Columbia, Vancouver, BC, V6T1Z4, Canada}}
\newcommand{\AFFbu}{\affiliation{Department of Physics, Boston University, Boston, MA 02215, USA}}
\newcommand{\AFFuci}{\affiliation{Department of Physics and Astronomy, University of California, Irvine, Irvine, CA 92697-4575, USA }}
\newcommand{\AFFcsu}{\affiliation{Department of Physics, California State University, Dominguez Hills, Carson, CA 90747, USA}}
\newcommand{\AFFcnm}{\affiliation{Department of Physics, Chonnam National University, Kwangju 500-757, Korea}}
\newcommand{\AFFduke}{\affiliation{Department of Physics, Duke University, Durham NC 27708, USA}}
\newcommand{\AFFfukuoka}{\affiliation{Junior College, Fukuoka Institute of Technology, Fukuoka, Fukuoka 811-0295, Japan}}
\newcommand{\AFFgifu}{\affiliation{Department of Physics, Gifu University, Gifu, Gifu 501-1193, Japan}}
\newcommand{\AFFgist}{\affiliation{GIST College, Gwangju Institute of Science and Technology, Gwangju 500-712, Korea}}
\newcommand{\AFFuh}{\affiliation{Department of Physics and Astronomy, University of Hawaii, Honolulu, HI 96822, USA}}
\newcommand{\AFFicl}{\affiliation{Department of Physics, Imperial College London , London, SW7 2AZ, United Kingdom }}
\newcommand{\AFFkek}{\affiliation{High Energy Accelerator Research Organization (KEK), Tsukuba, Ibaraki 305-0801, Japan }}
\newcommand{\AFFkobe}{\affiliation{Department of Physics, Kobe University, Kobe, Hyogo 657-8501, Japan}}
\newcommand{\AFFkyoto}{\affiliation{Department of Physics, Kyoto University, Kyoto, Kyoto 606-8502, Japan}}
\newcommand{\AFFliv}{\affiliation{Department of Physics, University of Liverpool, Liverpool, L69 7ZE, United Kingdom}}
\newcommand{\AFFmiyagi}{\affiliation{Department of Physics, Miyagi University of Education, Sendai, Miyagi 980-0845, Japan}}
\newcommand{\AFFnagoya}{\affiliation{Institute for Space-Earth Environmental Research, Nagoya University, Nagoya, Aichi 464-8602, Japan}}
\newcommand{\AFFkmi}{\affiliation{Kobayashi-Maskawa Institute for the Origin of Particles and the Universe, Nagoya University, Nagoya, Aichi 464-8602, Japan}}
\newcommand{\AFFpol}{\affiliation{National Centre For Nuclear Research, 02-093 Warsaw, Poland}}
\newcommand{\AFFsuny}{\affiliation{Department of Physics and Astronomy, State University of New York at Stony Brook, NY 11794-3800, USA}}
\newcommand{\AFFokayama}{\affiliation{Department of Physics, Okayama University, Okayama, Okayama 700-8530, Japan }}
\newcommand{\AFFosaka}{\affiliation{Department of Physics, Osaka University, Toyonaka, Osaka 560-0043, Japan}}
\newcommand{\AFFox}{\affiliation{Department of Physics, Oxford University, Oxford, OX1 3PU, United Kingdom}}
\newcommand{\AFFqmul}{\affiliation{School of Physics and Astronomy, Queen Mary University of London, London, E1 4NS, United Kingdom}}
\newcommand{\AFFregina}{\affiliation{Department of Physics, University of Regina, 3737 Wascana Parkway, Regina, SK, S4SOA2, Canada}}
\newcommand{\AFFseoul}{\affiliation{Department of Physics, Seoul National University, Seoul 151-742, Korea}}
\newcommand{\AFFsheff}{\affiliation{Department of Physics and Astronomy, University of Sheffield, S3 7RH, Sheffield, United Kingdom}}
\newcommand{\AFFshizuokasc}{\affiliation{Department of Informatics in
Social Welfare, Shizuoka University of Welfare, Yaizu, Shizuoka, 425-8611, Japan}}
\newcommand{\AFFstfc}{\affiliation{STFC, Rutherford Appleton Laboratory, Harwell Oxford, and Daresbury Laboratory, Warrington, OX11 0QX, United Kingdom}}
\newcommand{\AFFskk}{\affiliation{Department of Physics, Sungkyunkwan University, Suwon 440-746, Korea}}
\newcommand{\AFFtokyo}{\affiliation{The University of Tokyo, Bunkyo, Tokyo 113-0033, Japan }}
\newcommand{\AFFtodai}{\affiliation{Department of Physics, University of Tokyo, Bunkyo, Tokyo 113-0033, Japan }}
\newcommand{\AFFtit}{\affiliation{Department of Physics,Tokyo Institute of Technology, Meguro, Tokyo 152-8551, Japan }}
\newcommand{\AFFtus}{\affiliation{Department of Physics, Faculty of Science and Technology, Tokyo University of Science, Noda, Chiba 278-8510, Japan }}
\newcommand{\AFFtoronto}{\affiliation{Department of Physics, University of Toronto, ON, M5S 1A7, Canada }}
\newcommand{\AFFtriumf}{\affiliation{TRIUMF, 4004 Wesbrook Mall, Vancouver, BC, V6T2A3, Canada }}
\newcommand{\AFFtokai}{\affiliation{Department of Physics, Tokai University, Hiratsuka, Kanagawa 259-1292, Japan}}
\newcommand{\AFFtsinghua}{\affiliation{Department of Engineering Physics, Tsinghua University, Beijing, 100084, China}}
\newcommand{\AFFynu}{\affiliation{Faculty of Engineering, Yokohama National University, Yokohama, Kanagawa, 240-8501, Japan}}
\newcommand{\AFFllr}{\affiliation{Ecole Polytechnique, IN2P3-CNRS, Laboratoire Leprince-Ringuet, F-91120 Palaiseau, France }}
\newcommand{\AFFbari}{\affiliation{ Dipartimento Interuniversitario di Fisica, INFN Sezione di Bari and Universit\`a e Politecnico di Bari, I-70125, Bari, Italy}}
\newcommand{\AFFnapoli}{\affiliation{Dipartimento di Fisica, INFN Sezione di Napoli and Universit\`a di Napoli, I-80126, Napoli, Italy}}
\newcommand{\AFFroma}{\affiliation{INFN Sezione di Roma and Universit\`a di Roma ``La Sapienza'', I-00185, Roma, Italy}}
\newcommand{\AFFpadova}{\affiliation{Dipartimento di Fisica, INFN Sezione di Padova and Universit\`a di Padova, I-35131, Padova, Italy}}
\newcommand{\AFFkeio}{\affiliation{Department of Physics, Keio University, Yokohama, Kanagawa, 223-8522, Japan}}
\newcommand{\AFFwinnipeg}{\affiliation{Department of Physics, University of Winnipeg, MB R3J 3L8, Canada }}
\newcommand{\AFFkcl}{\affiliation{Department of Physics, King's College London, London, WC2R 2LS, UK }}
\newcommand{\AFFwarwick}{\affiliation{Department of Physics, University of Warwick, Coventry, CV4 7AL, UK }}
\newcommand{\AFFral}{\affiliation{Rutherford Appleton Laboratory, Harwell, Oxford, OX11 0QX, UK }}

\AFFicrr
\AFFkashiwa
\AFFmad
\AFFbu
\AFFuci
\AFFcsu
\AFFcnm
\AFFduke
\AFFllr
\AFFfukuoka
\AFFgifu
\AFFgist
\AFFuh
\AFFicl
\AFFbari
\AFFnapoli
\AFFpadova
\AFFroma
\AFFkcl
\AFFkeio
\AFFkek
\AFFkobe
\AFFkyoto
\AFFliv
\AFFmiyagi
\AFFnagoya
\AFFkmi
\AFFpol
\AFFsuny
\AFFokayama
\AFFosaka
\AFFox
\AFFral
\AFFseoul
\AFFsheff
\AFFshizuokasc
\AFFstfc
\AFFskk
\AFFtokai
\AFFtokyo
\AFFtodai
\AFFipmu
\AFFtit
\AFFtus
\AFFtoronto
\AFFtriumf
\AFFtsinghua
\AFFwarwick
\AFFwinnipeg
\AFFynu

\author{M.~Tanaka}
\AFFtit
\author{K.~Abe}
\AFFicrr
\AFFipmu
\author{C.~Bronner}
\AFFicrr
\author{Y.~Hayato}
\AFFicrr
\AFFipmu
\author{M.~Ikeda}
\author{S.~Imaizumi}
\AFFicrr
\author{H.~Ito}
\AFFicrr 
\author{J.~Kameda}
\AFFicrr
\AFFipmu
\author{Y.~Kataoka}
\AFFicrr
\author{Y.~Kato}
\AFFicrr
% ...


\collaboration{The Super-Kamiokande Collaboration}
\noaffiliation

\date{\today}

\begin{abstract}
A search for proton decay into three charged leptons has been performed by using 0.37\,Mton$\cdot$years of data collected in Super-Kamiokande.
All possible combinations of electrons, muons and their anti-particles consistent with charge conservation were considered as decay modes.
No significant excess of events has been found over the background, and lower limits on the proton lifetime divided by the branching ratio have been obtained.
The limits range between $9.2\times10^{33}$ to $3.4\times10^{34}$ years at 90\% confidence level, improving by more than an order of magnitude upon limits from previous experiments.
A first limit has been set for the $p\rightarrow\mu^-e^+e^+$ mode.

\end{abstract}

\maketitle


\section{Introduction}
The Standard Model of elementary particles describes strong, weak and electromagnetic interactions based on gauge symmetries.
Grand Unified Theories (GUTs) \cite{gut_theory} can unify three interactions in the Standard Model in a single gauge group with one coupling constant.
In most GUTs, grand unification is predicted typically at  $10^{15-16}$\,GeV which is unreachable by accelerators, whereas the effects of the grand unification might be detected through rare phenomena beyond the Standard Model.
The most promising such phenomenon is violation of baryon number, and the most sought after signature is proton decay \cite{gut_pdk}.
The Super-Kamiokande (SK) experiment has been leading the search for proton decay and has set the most stringent limits on the lifetime for various channels predicted by GUT models.
For example, the $p \rightarrow e^{+}  \pi^{0}$ and $p \rightarrow \overline{\nu}  K^{+}$ are favored decay modes in non-supersymmetric and TeV-scale supersymmetric GUT models, respectively, yet no significant signal was observed, excluding proton lifetimes up to $10^{34}$ years \cite{pdk_sk_epi, pdk_sk_nuk}.
The simplest unification model, minimal SU(5) \cite{su5_theory}, has been ruled out by SK and earlier experiments.
Other nucleon decay channels motivated by unification, such as the charged antilepton plus meson channels were searched for in SK recently; no proton decay signal was found \cite{pdk_sk_antilepton_meson}.

However, baryon number may be violated irrespective of GUTs and is expected in many scenarios beyond the Standard Model \cite{proton_decay}. 
If the usual lower dimensional operators responsible for $p \rightarrow e^{+}  \pi^{0}$ etc. are suppressed, then different proton decay channels can dominate from higher dimensional operators. 
This can naturally occur when considering lepton flavor symmetries \cite{p_charged_leptons}. 
In this case, trilepton nucleon decay channels such as $p\rightarrow \mu^-e^+e^+$ or $p\rightarrow e^-\mu^+\mu^+$  can be dominant. 
As they are generated from effective dimension $d=10$ operators, these processes probe scales of around 100\,TeV. 
A minimal model for this based on leptoquarks has been put forward in Ref.~\cite{p_charged_leptons}, which also suggested that such processes could be connected to the recent anomalies observed in B-meson decays.

Some of these decay modes were already searched for by the Irvine-Michigan-Brookhaven-3 (IMB-3) \cite{IMB_proton_decay} and Harvard-Purdue-Wisconsin (HPW) \cite{HPW_proton_decay} experiments.
The data were consistent with the expected background in both experiments and no significant signal was confirmed.
Lifetime limits were set to be $10^{30}$-$10^{32}$ years for each decay mode. 
SK can significantly extend the search of the previous experiments.

\section{The Super-Kamiokande Experiment}
SK is the largest pure water Cherenkov detector, located at the Kamioka mine in Gifu prefecture, Japan.
The SK detector consists of a stainless steel tank (39.3\,m diameter, 41.4\,m height), 50\,kton of ultra pure water and photomultiplier tubes (PMTs).
In order to reduce cosmic ray muon background, the detector is located 1,000\,m under the peak of Mt. Ikenoyama (2,700\,m water equivalent).
The water tank is optically separated into two concentric cylindrical volumes by the support frames equipped with inward-facing 20-inch and outward-facing 8-inch PMTs.
The inner volume is 33.8\,m in diameter and 36.2\,m in height, and is called the Inner Detector (ID).
The ID contains 32\,kton of water and monitored by 11,129 inward facing PMTs (about half for certain period as explained later).
Outside of the ID is a 2\,m thick water called the Outer Detector (OD).
The OD is monitored by outward-facing 1,885 PMTs and mainly serves as an active cosmic ray muon veto as well as a shield against gamma rays from surrounding rock.
20-inch and 8-inch PMTs are uniformly mounted on the ID and OD surfaces, respectively. 
The details of SK detector are described elsewhere \cite{SK_detector, SK_detector2}.

The analysis in this paper uses data taken with four different detector configurations.
SK-I started in 1996 and stopped in 2001 for maintenance.
SK-II was operated from 2002 to 2005 with about half the number of ID PMTs compared to SK-I due to the accident during the maintenance work after SK-I.
In order to avoid further such accidents, the PMTs were protected by covers made of fiber reinforced plastic and acrylic for the photocathode starting from SK-II.
SK-III started in 2006 and stopped in 2008.
The number of PMTs in SK-III was recovered to almost the same number as SK-I.
Readout electronics and data acquisition system were upgraded for the SK-IV period.
SK-IV continued from 2008 and ended for the upgrade of SK in May 2018.
Photocoverage of the ID is 19\% in SK-II and 40\% in other periods.

In this analysis, all the detected particles must be fully contained (FC) in the ID with the reconstructed vertex inside the fiducial volume (FV).
Such events are selected by preselection cuts \cite{pdk_sk_epi, pdk_sk_nuk, pdk_sk_antilepton_meson}.
The FV is defined as the volume 2\,m inside the top, bottom and barrel walls of the ID and corresponds to 22.5\,kton mass.
Contamination of non-neutrino background events due to cosmic muons is negligible after the preselection cuts.
Data around 1\,ms of the expected neutrino beam arrival timing from the Japan Proton Accelerator Research Complex (J-PARC), which has a repetition rate of 2.48 s, have been removed.
All the available data from SK-I to SK-IV are used in this analysis.
We use the data of 372.6\,kton$\cdot$years in total by summing up 91.7, 49.2, 31.9 and 199.8\,kton$\cdot$years of SK-I, II, III and IV data, respectively.


\section{Simulation}
H$_2$O molecules are the sources of proton decay in SK searches, with 2 protons from hydrogen and 8 protons from oxygen.
In the simulation, only a uniform phase space is assumed for kinematics of outgoing charged leptons and any additional correlations are not taken into account.
The protons in hydrogen (free protons) have an initial mass and momentum of 938.27\,MeV/c$^2$ and 0\,MeV/c, respectively.
On the other hand, the protons in oxygen (bound protons) interact with other nucleons and have some initial momentum.
In the simulation, three nuclear effects are taken into account: nuclear binding energies in $^{16}$O, Fermi motion, and correlated decay.
Two nuclear binding energies ($p$-state and $s$-state) are accounted for with Gaussian spreads and are subtracted from the initial proton mass \cite{pdk_sk_antilepton_meson}.
Fermi motion is estimated based on electron-$^{12}$C scattering data \cite{exp_fermi_motion}.
The bound proton sometimes correlates with the surrounding nucleons during its decay.
This effect is predicted to occur with 10\% probability and produces a broad distribution at lower mass in proton mass distribution \cite{correlated_decay}.

For the background, only atmospheric neutrino events are considered since other non-neutrino backgrounds are negligibly small.
The simulation of this sample consists of three steps: neutrino production in the atmosphere (neutrino flux prediction), neutrino interaction with water and particle tracking in the detector.
The flux of atmospheric neutrino is calculated by the model of M. Honda {\it et al.} \cite{neutrino_flux, neutrino_flux_2}.
The interactions of atmospheric neutrinos with hydrogen or oxygen nuclei in water are simulated by the NEUT program \cite{neutrino_interaction}.
This simulation covers a wide neutrino energy range from several tens of MeV to hundreds of TeV.
Hadrons generated by neutrino interactions in the oxygen nucleus often cause secondary interactions within the nucleus.
The interactions of pions, kaons, etas and nucleons in the target nucleus are simulated in NEUT by using a cascade model \cite{neutrino_interaction, neut_fsi}.
Simulated data samples equivalent to 500 years of detector exposure are generated for each SK period.

The particle propagation, Cherenkov radiation, propagation of Cherenkov photons in water and PMTs, and electronics response are simulated by a Geant3 based package \cite{geant3} with custom modifications for use in SK, such as pion interactions in water and wavelength-dependent water transparency.

The simulation scheme for the signal sample is the same as for the other recent SK nucleon decay searches \cite{pdk_sk_epi, pdk_sk_nuk, pdk_sk_antilepton_meson}, and the latest SK oscillation analysis \cite{latest_oscillation} for background sample.
The oscillation parameters are taken from the latest atmospheric
neutrino oscillation analysis \cite{latest_oscillation}.

\section{Event reconstruction}
Events with charged particles are reconstructed by using charge and time information of the hit PMTs.
A reconstruction scheme consists of vertex fitting, ring counting, particle identification (PID), momentum reconstruction, Michel electron search and neutron tagging.
The reconstruction method is almost the same as the one used in other recent nucleon decay searches \cite{pdk_sk_epi,pdk_sk_nuk, pdk_sk_antilepton_meson} or oscillation studies \cite{latest_oscillation} in SK.
Some updates for charge separation and neutron tagging improve the sensitivity of this search.

In the first step, the vertex is reconstructed by assuming that Cherenkov light comes from one point at the same time.
Then the ring edge and the direction of the ring are estimated.
Finally, the vertex is reconstructed more precisely by assuming that photons are emitted along the track of the charged particle.

Cherenkov rings are then reconstructed by using the pattern recognition algorithm known as the Hough transformation \cite{Hough}.
Ring candidates are evaluated by a likelihood method to determine if they are true or fake.
In case more than one rings (multi-ring) are identified, the contribution of each ring to the detected photoelectrons in each PMT is estimated.
The opening angle of the ring can be calculated from the reconstructed vertex position and the edge of the ring.
The final stage of ring counting discards the candidate rings mostly caused by multiple Coulomb scattering of charged particles by a ring's angle relative to other rings and by visible energy.

Reconstructed rings can be classified as electron-like ($e$-like) or muon-like ($\mu$-like) by using the pattern of PMT hits.
Cherenkov rings of muons tend to have clear ring edges.
In contrast, Cherenkov rings of electrons tend to be relatively diffuse due to electromagnetic showers and scattering.
Expected PMT charge patterns for electrons and muons are compared with observed hit patterns using a likelihood function.  
Information about the opening angle is included in the likelihood function.
With the emission of Cherenkov rings, gamma rays are usually identified as $e$-like and charged pions as $\mu$-like.

The momentum of the charged particle is reconstructed from the total number of photoelectrons in a 70\,degree cone around the ring, which is corrected by using a conversion table depending on the particle type.
We correct for time drift of the gain, which varies according to the year of PMT manufacture.
For the multi-ring case, the expected charge distribution for each ring is calculated.
Then the momentum is assigned to each ring according to the fraction of expected charge.
For this charge separation algorithm, the expected charge in the backward direction of a $\mu$-like ring was tuned to reproduce the data. 
The precision of total mass reconstruction was improved with this tuning, especially for the $p\rightarrow\mu^+\mu^+\mu^-$ events.
The energy scale of the detector is checked precisely by using Michel electrons, stopping muons and neutral pion samples \cite{latest_oscillation}.

Michel electrons are tagged by searching for PMT hit clusters after the primary event.
Since about 20\% of $\mu^-$ are captured by nuclei and do not emit a decay electron, the tagging efficiency for $\mu^-$ is lower than that for $\mu^+$.

Free neutrons traveling in water are thermalized and captured by oxygen or hydrogen nuclei. 
Neutrons are predominantly captured by the interaction $n+p\rightarrow d+\gamma$ (2.2\,MeV).
This 2.2 MeV gamma ray signal is searched for to identify the neutron (neutron tagging).
The capture signal occurs a few hundred microseconds after the initial neutrino interaction signal, and the tagging was possible only with the improved electronics introduced in SK-IV.
The performance of the neutron tagging was recently improved by lowering the neutron tagging threshold (the maximum number of hit PMTs within 10\,ns sliding time window) thanks to an additional cut on the continuous dark noise hits after one initial dark noise pulse and new parameters in the neural network.
The tagging efficiency was improved from 22\% \cite{ntag} to 25\%.

\section{Event selection \label{section_event_selection}}
The following selection criteria are applied to separate proton decay signals from atmospheric neutrino background events.
The selection criteria resemble those of other recent SK nucleon decay searches \cite{pdk_sk_epi,pdk_sk_antilepton_meson}.
The same selections are applied to the data and MC simulation (both proton decay signal and atmospheric neutrino background).

\begin{description}
\item[C1] 
There must be three reconstructed Cherenkov rings.
\item[C2] 
PID of Cherenkov rings must be consistent with the decay mode.
For example, there must be one $e$-like ring and two $\mu$-like rings for the $p\rightarrow e^+\mu^+\mu^-$ and $p\rightarrow e^-\mu^+\mu^+$ decay modes.
SK cannot tell the charge of the final state lepton, so that cuts and backgrounds for the $p\rightarrow \mu^+e^+e^-$ and $p\rightarrow \mu^-e^+e^+$ ($p\rightarrow e^+\mu^+\mu^-$ and $p\rightarrow e^-\mu^+\mu^+$) are essentially the same.
\item[C3] 
Numbers of decay electrons should be 0 for the $p\rightarrow e^+e^+e^-$, 1 for the $p\rightarrow \mu^+e^+e^-$ ($p\rightarrow \mu^-e^+e^+$), 2 for the $p\rightarrow e^+\mu^+\mu^-$ ($p\rightarrow e^-\mu^+\mu^+$), and 2 or 3 for the $p\rightarrow \mu^+\mu^+\mu^-$ decay modes.
\item[C4] 
Total mass ($M_{\rm tot}$) and momentum ($P_{\rm tot}$) of 3-ring events should satisfy $800<M_{\rm tot}<1050$\,MeV/c$^2$ and $P_{\rm tot}<250$\,MeV/c. 
In case of the $p\rightarrow\mu^+e^+e^-$ ($p\rightarrow\mu^-e^+e^+$) mode, one additional cut is used. 
The invariant mass of two $e$-like rings events should be above 185\,MeV/c$^2$.
\item[C5]
There should be no tagged neutron (only for SK-IV).
\end{description}

\begin{figure*}[htbp]
	\centering
	\includegraphics[width=55mm]{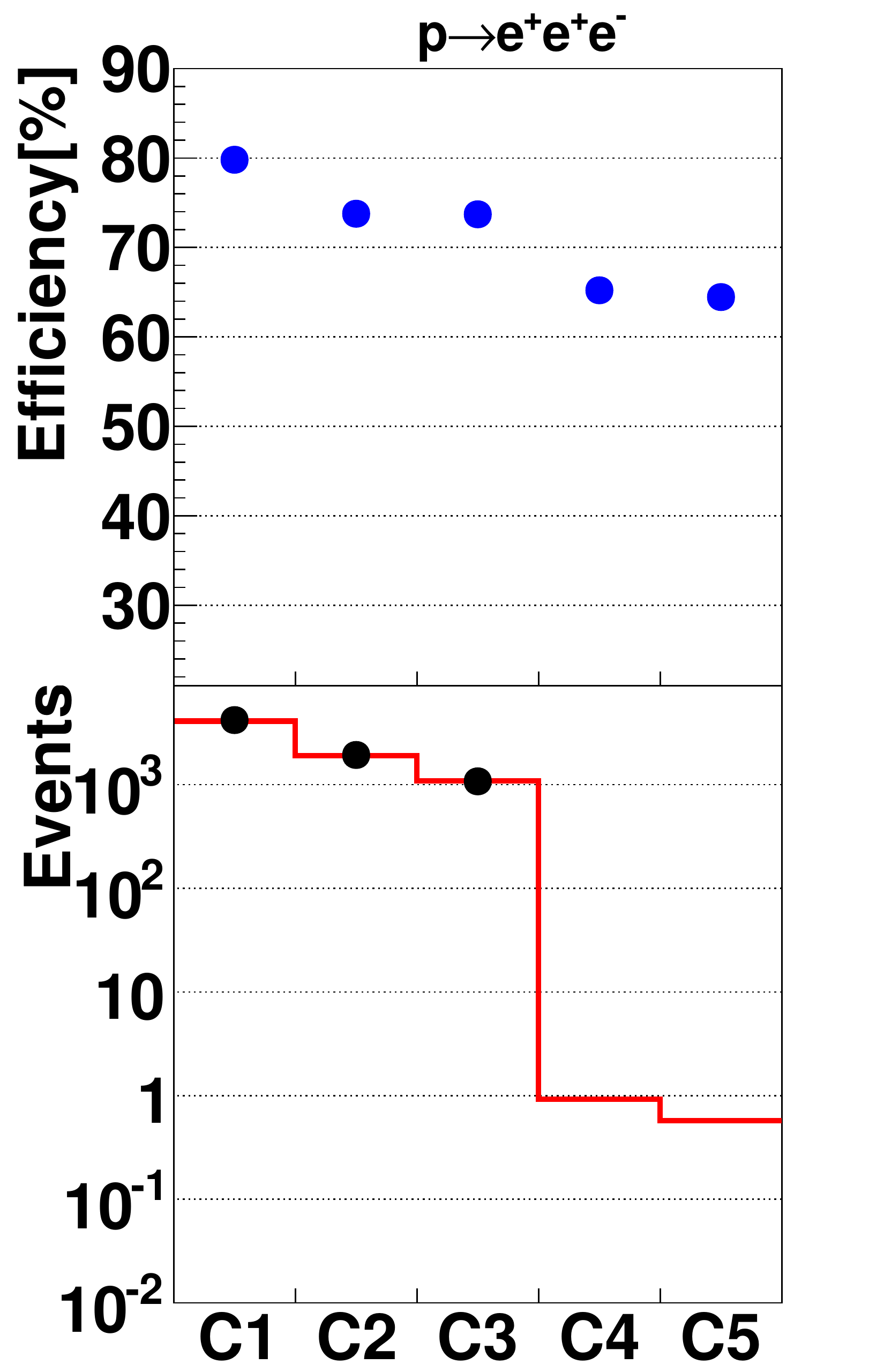}
	\includegraphics[width=55mm]{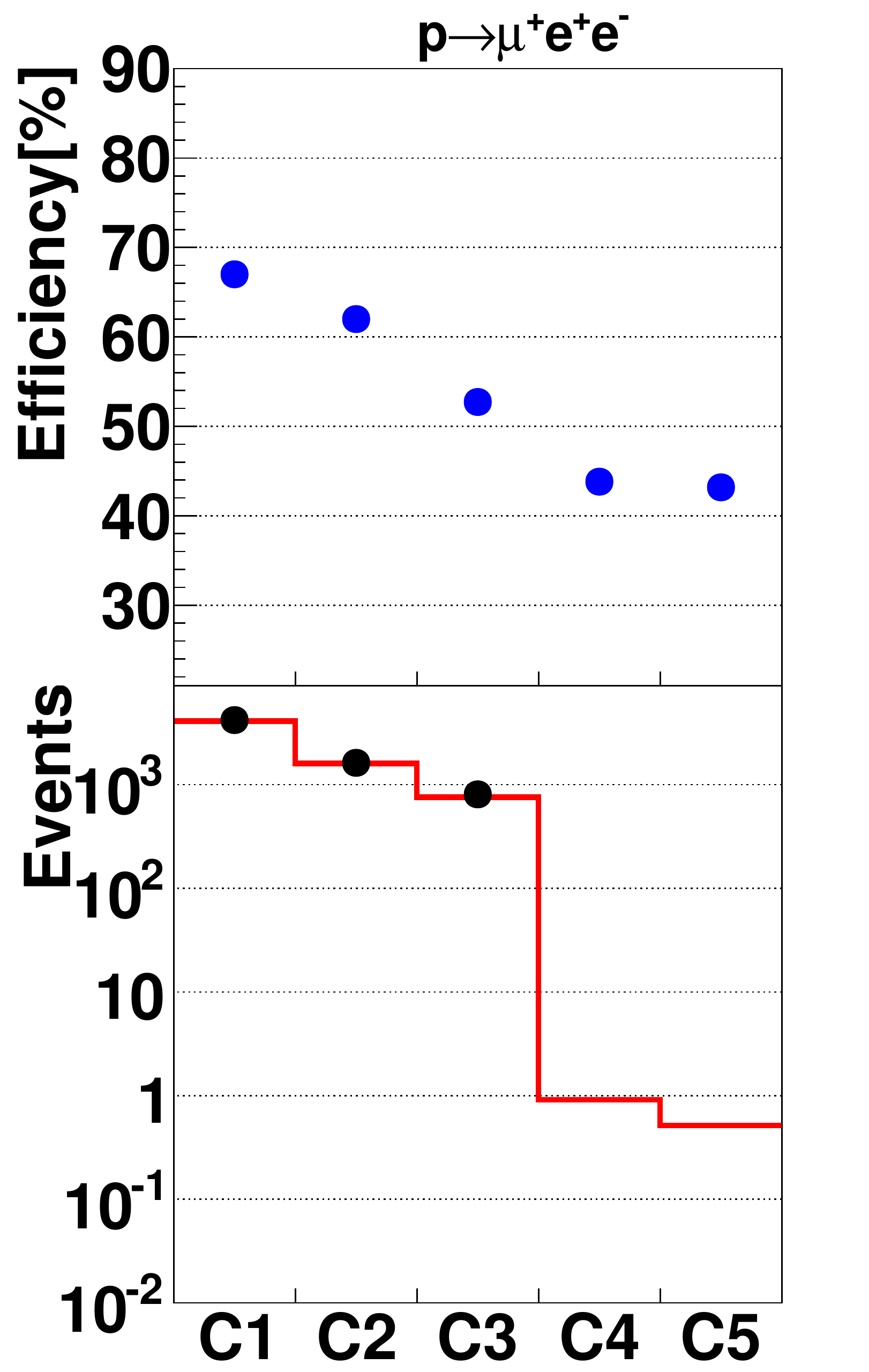}
	\includegraphics[width=55mm]{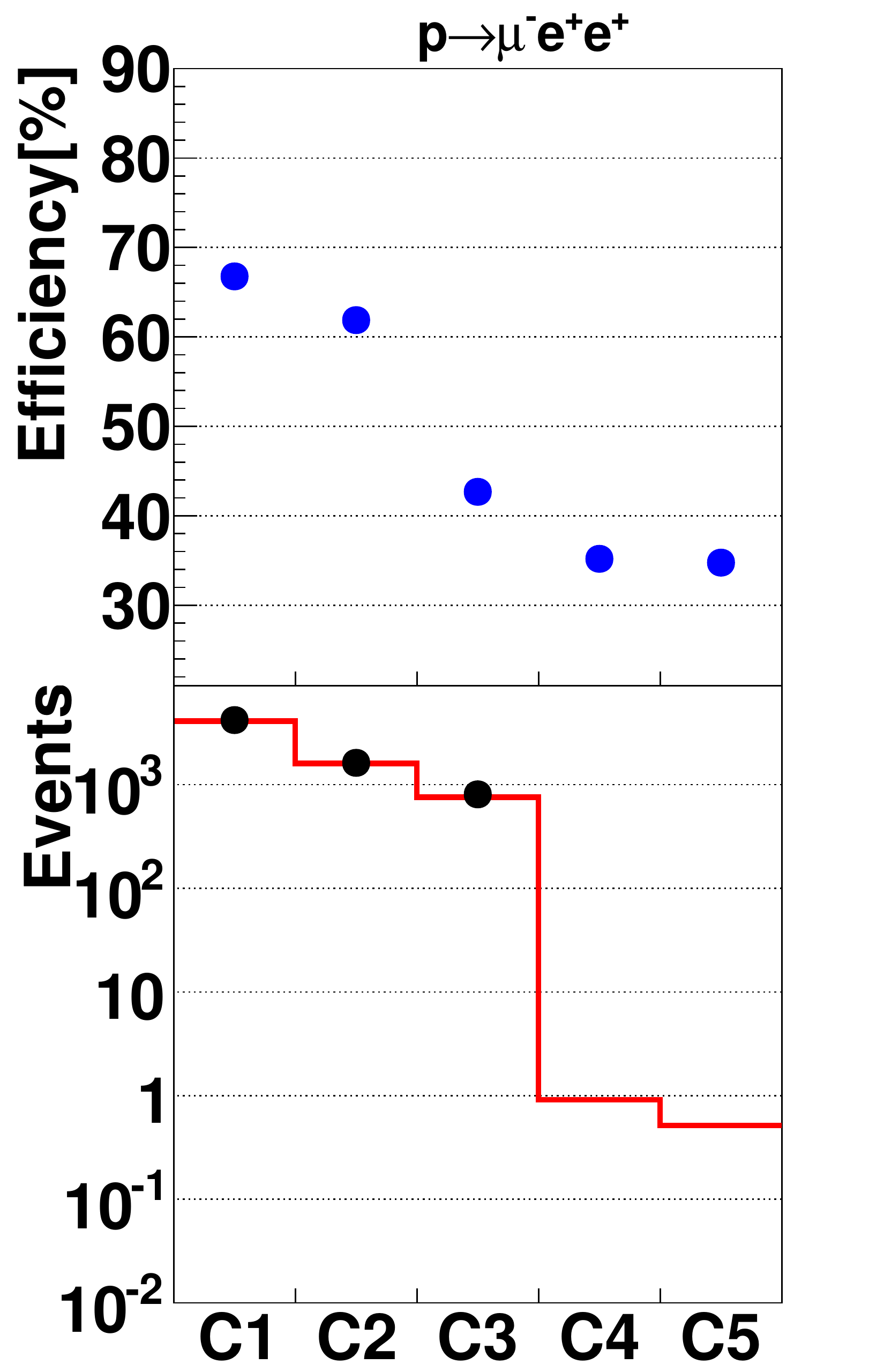}
	\includegraphics[width=55mm]{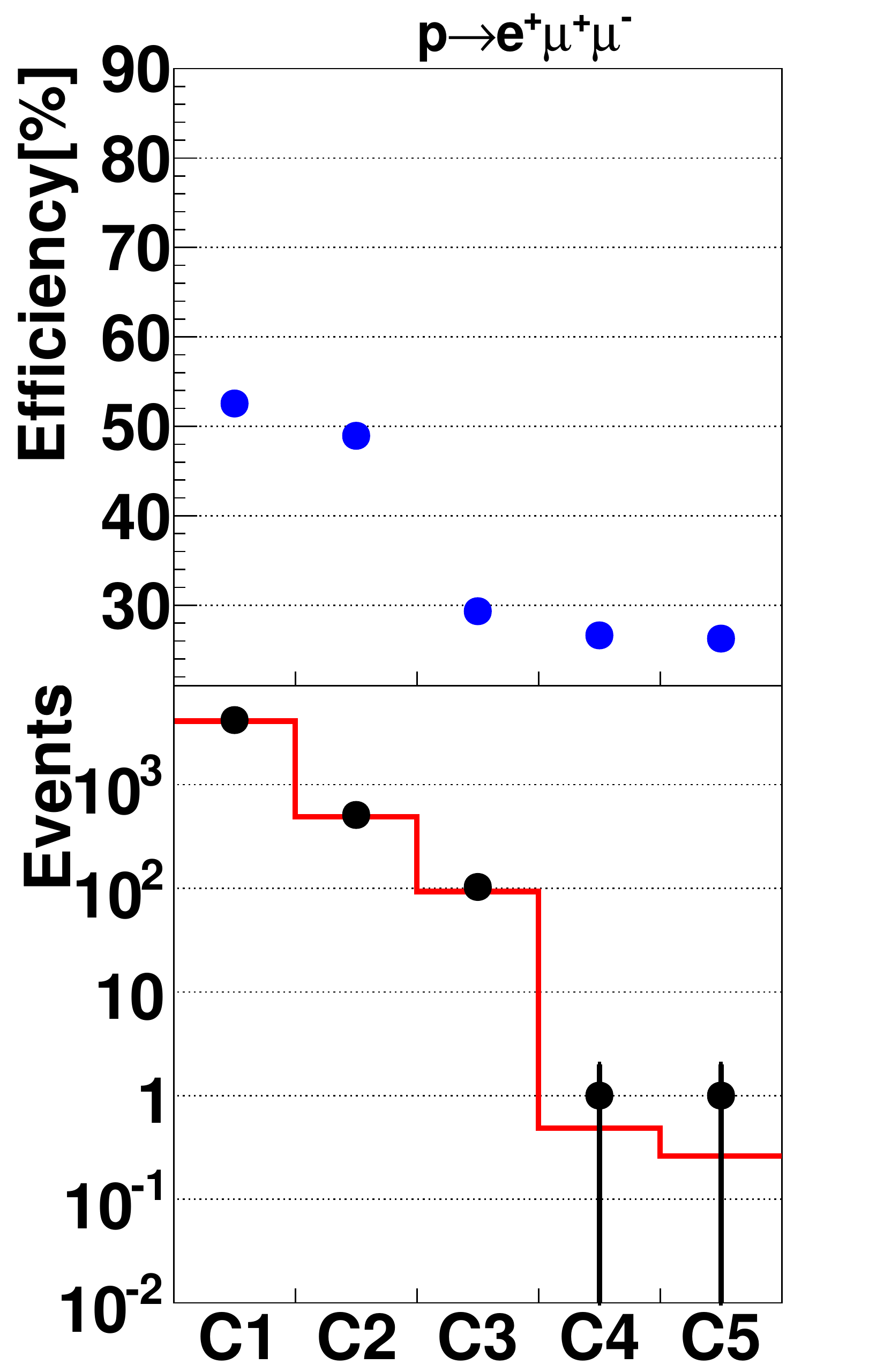}
	\includegraphics[width=55mm]{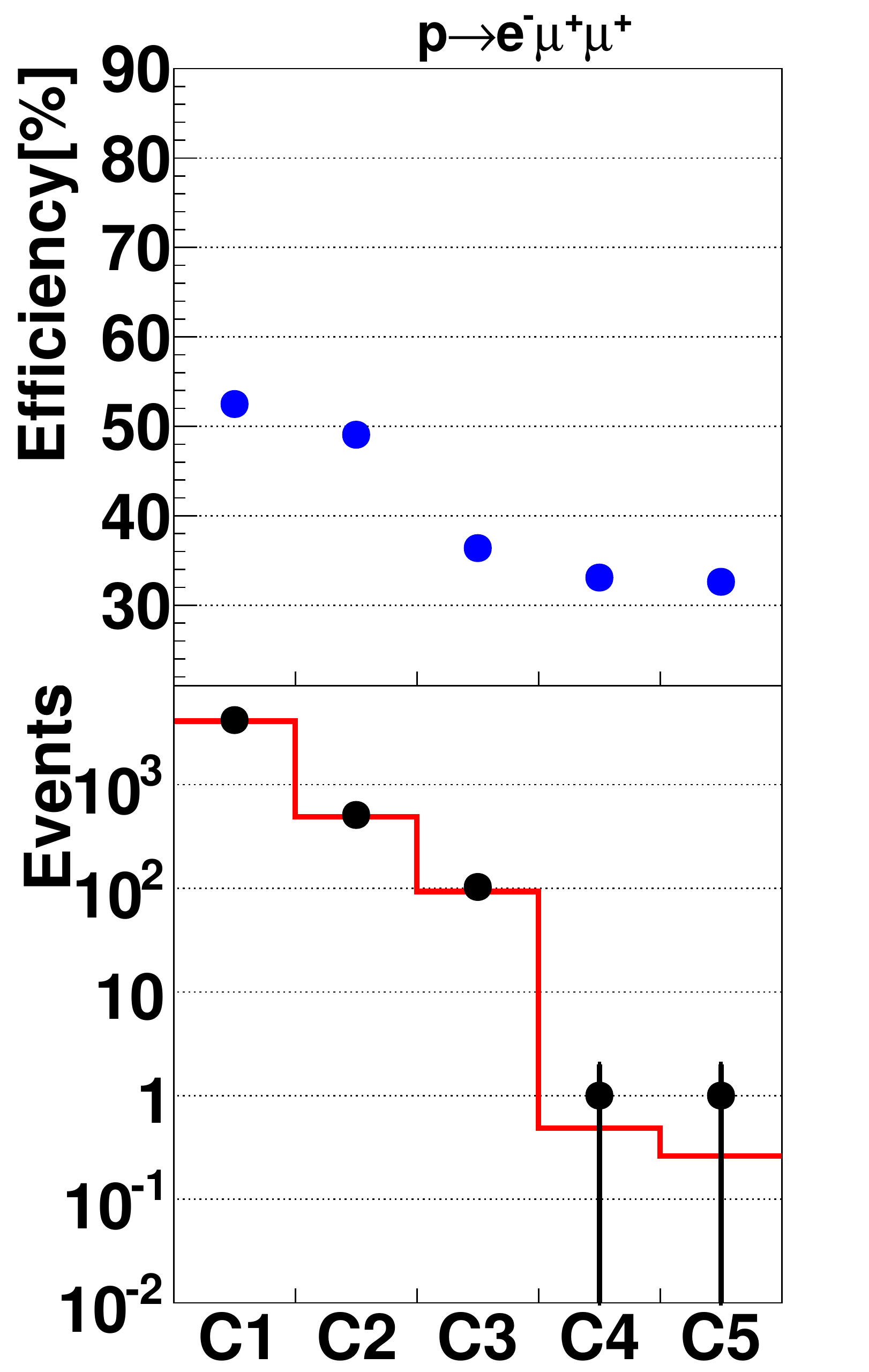}
	\includegraphics[width=55mm]{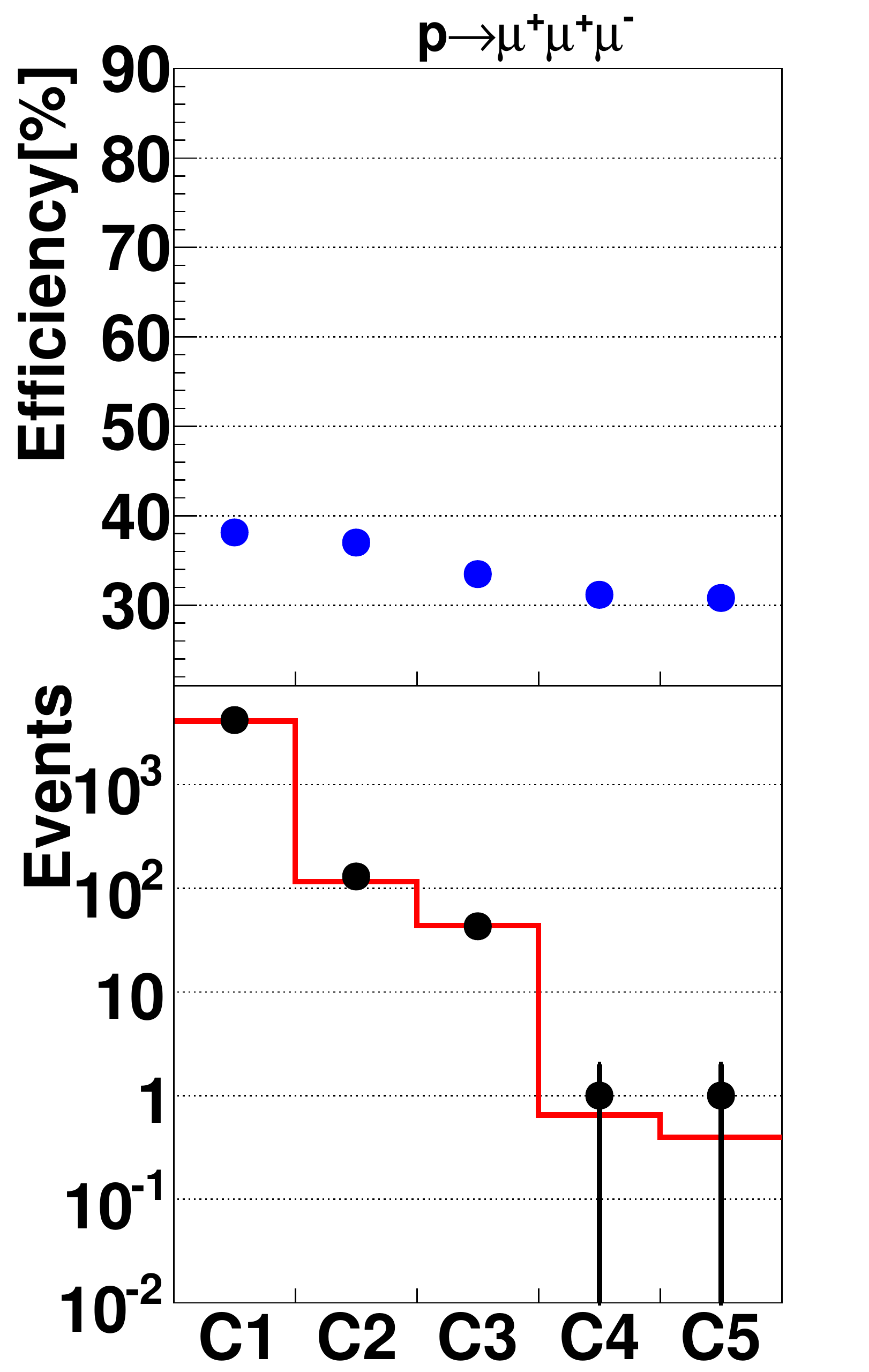}
	\caption{Signal efficiency (top, blue dots), data (bottom, black dots) and expected background events (bottom, red line) at each cut step in each mode. In upper column, from left to right for the $p\rightarrow e^+e^+e^-$, $p\rightarrow \mu^+e^+e^-$ and $p\rightarrow \mu^-e^+e^+$ modes, respectively. In lower column, from left to right for the $p\rightarrow e^+\mu^+\mu^-$, $p\rightarrow e^-\mu^+\mu^+$ and $p\rightarrow \mu^+\mu^+\mu^-$ modes, respectively. The background MC is normalized by livetime.  SK-I-IV are combined in signal MC, background MC and data. 
	Note that both the data and background MC plots for the $p\rightarrow\mu^+e^+e^-$ and $p\rightarrow \mu^-e^+e^+$ modes are the same, since SK cannot identify the change sign of the leptons.  
	For the same reason, the data and background MC plots are the same for the $p\rightarrow e^+\mu^+\mu^-$ and $p\rightarrow e^-\mu^+\mu^+$ modes as well. }
	\label{fig_cutflow}
\end{figure*}

These cuts are applied to reduce atmospheric neutrino background, mainly deep inelastic scattering (DIS), based on the kinematics of the outgoing charged particles.
Signal selection efficiencies of the C1 cut are lower for decay modes with more muons as shown in FIG.\ref{fig_cutflow}.
This is due to the higher Cherenkov threshold for muons compared to that for electrons.
More than 90\% of atmospheric neutrino background events are rejected by requiring three rings.

The C3 cut requires a number of Michel electrons depending on the number of muons in the final state.
For the $p\rightarrow \mu^+\mu^+\mu^-$ mode, 2 or 3 decay electrons are required to keep a good signal efficiency.
The signal efficiency of this cut depends on the charge and number of muons as the tagging efficiency of Michel electrons for $\mu^+$ is higher than that for $\mu^-$.
For example, efficiency for $p\rightarrow \mu^+e^+e^-$ is higher than that for the $p\rightarrow \mu^-e^+e^+$.

After the C3 cut is applied, the main background for the $p\rightarrow \mu^+e^+e^-$ mode is $\nu_{\mu}$ charged-current (CC) $\pi^0$ production events, in which a $\pi^0$ decays to two gamma rays and is identified as two $e$-like rings.
Such background can be reduced by requiring the invariant total mass of two $e$-like rings to be different from the $\pi^{0}$ mass (C4 cut for one-muon mode).
CC $\pi^0$ production events are the dominant background to the $p\rightarrow e^+e^+e^-$ as well (incoming neutrino is $\nu_{e}$ in this case).
However, since the background rejection is not beneficial to the proton decay signal efficiency, a cut on the invariant mass for two $e$-like rings is not applied for the $p\rightarrow e^+e^+e^-$ mode.

Total mass and momentum cuts (C4) are the most effective cut in this analysis.
They require that the kinematics of three charged particles is consistent with that from proton decay signals, i.e., their invariant mass should be consistent with the proton mass and the total momentum should be below the upper limit of Fermi motion of protons in oxygen nuclei (the momentum of the proton is 0 for free protons).
The lower and higher tails of the proton mass and momentum distributions become larger due to effects of correlated decay.

The probability for a neutron to be generated by de-excitation of a nucleus after proton decay is estimated to be a few percent \cite{ejiri}. 
On the other hand, neutrons are often generated in the background process, dominated by DIS interactions of atmospheric neutrinos in water.
By applying the C5 cut, about half of the background events are rejected while more than 90\% of the signal events are kept.

\begin{figure*}[htbp]
	\centering
	\includegraphics[width=160mm]{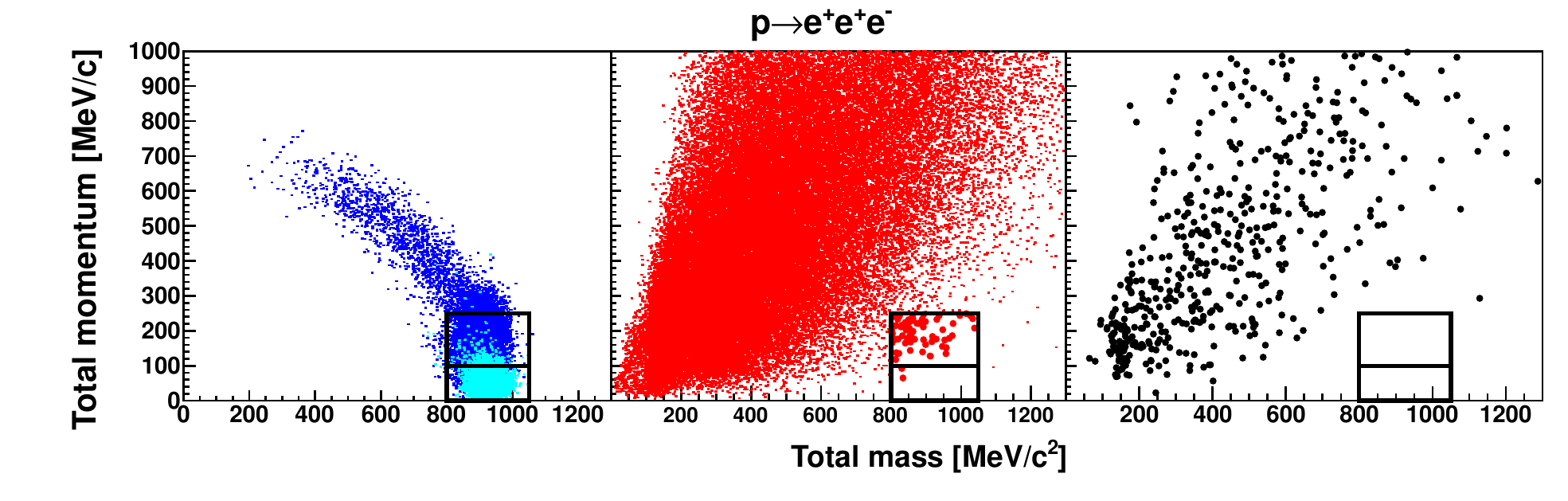}
	\includegraphics[width=160mm]{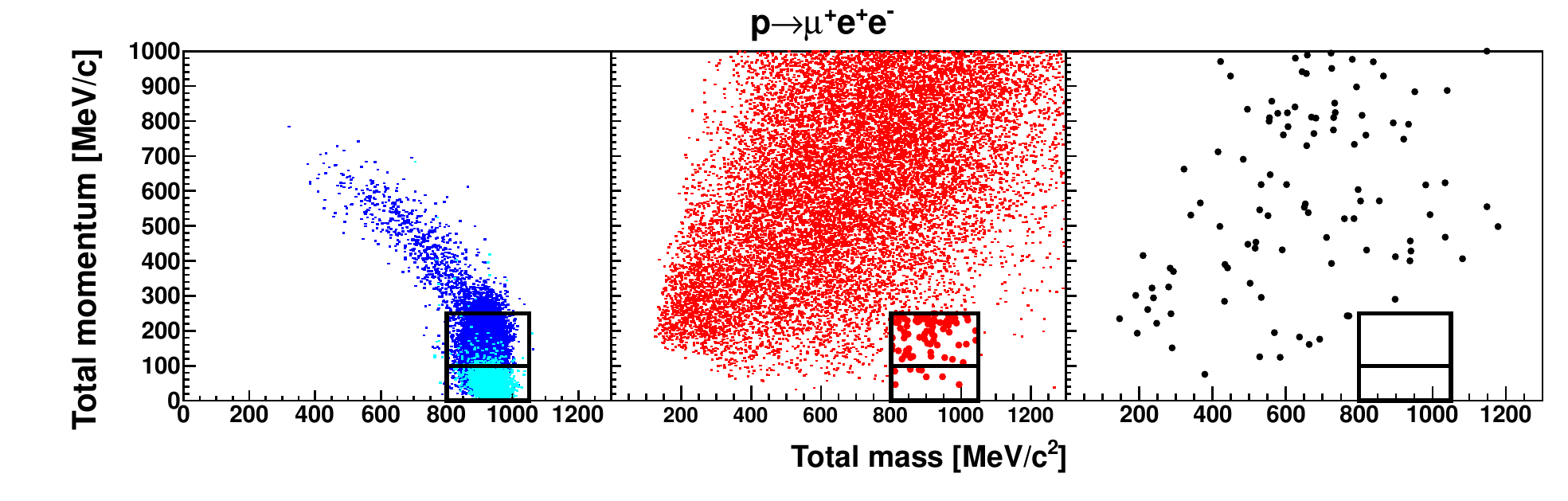}
	\includegraphics[width=160mm]{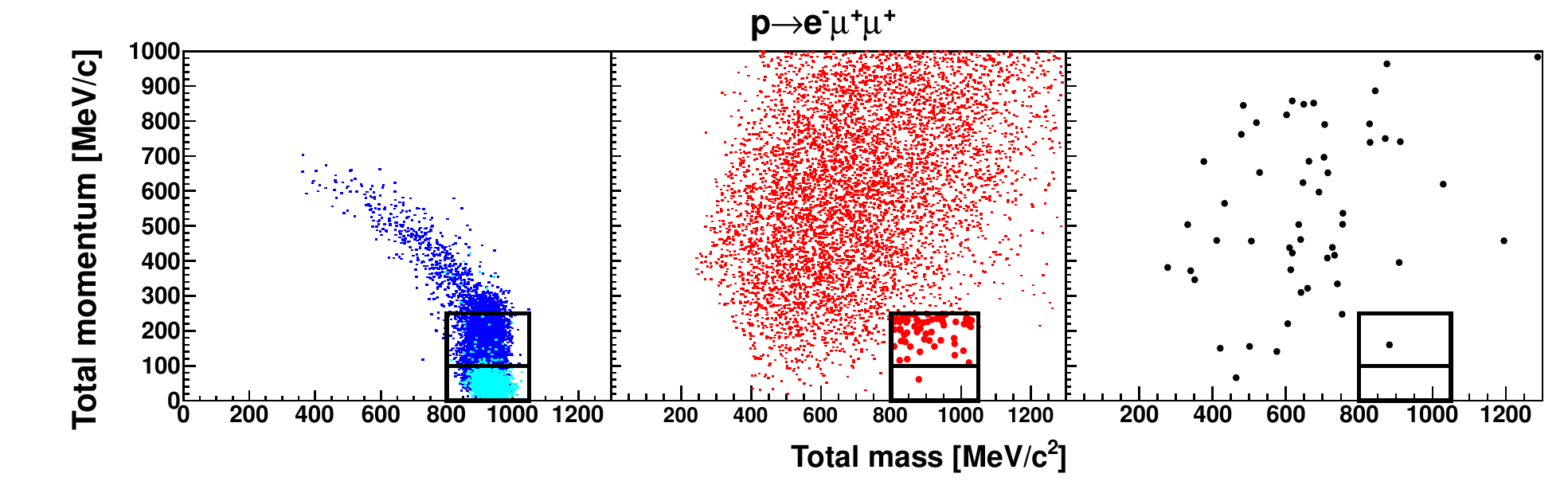}
	\includegraphics[width=160mm]{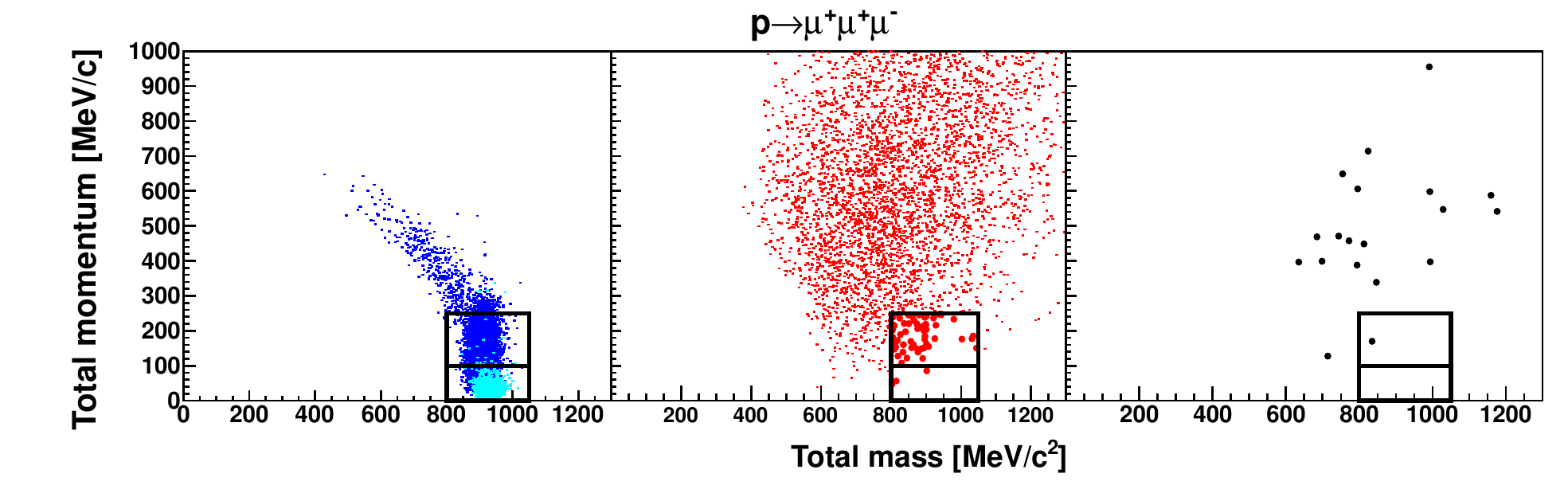}
	\caption{Two dimensional plots of total mass and momentum for signal (left), background (center) and measured data (right) after the selections C1-C3 and C5 are applied. 
	From the top to the bottom, for the $p\rightarrow e^+e^+e^-$, $p\rightarrow \mu^+e^+e^-$, $p\rightarrow e^-\mu^+\mu^+$ and $p\rightarrow \mu^+\mu^+\mu^-$ modes, respectively. 
	Light blue shows free proton and dark blue shows bound proton in the signal plot. 
	Two black squares show the lower and upper signal boxes. 
	The dot size is enlarged in the signal box only for background.
	SK-I-IV are combined in signal MC, background MC ($4\times500$ years), and data.
For the $p\rightarrow\mu^-e^+e^+$ mode the number of signal MC points is lower than that of $p\rightarrow\mu^+e^+e^-$ by 19\%, due to the different effective lifetimes (and therefore different decay electron production probabilities) of the $\mu^{-}$ and $\mu^{+}$ in water.
Similarly, the signal MC for $p\rightarrow e^+\mu^+\mu^-$ has 20\% fewer events than that for the $p\rightarrow e^-\mu^+\mu^+$ mode.
As in FIG.\ref{fig_cutflow}, the data and background MC figures are the same for modes that differ only by the charge sign of the leptons.
	}
	\label{fig_2D}
\end{figure*}

The number of events (data), the signal efficiency and the expected background along with the event rates after selection for each proton decay mode are shown in FIG.\ref{fig_cutflow}.
Scatter plots of total mass and momentum for signal and background MC are shown in FIG.\ref{fig_2D}.
Two signal boxes are defined: a lower signal box ($P_{\rm tot}<100$\,MeV/c) and an upper signal box ($100<P_{\rm tot}<250$\,MeV/c).
The lower signal box is almost background free and is dominated by free protons, while the proton bound in the oxygen nucleus is dominant in the upper signal box.
Signal efficiency and expected background events after all selections are summarized in TABLE \ref{table_selection_summary}.
Fractions of each neutrino interaction mode in the remaining background events are summarized in TABLE \ref{table_bkg_fraction}.
The dominant background is single or multi pion production events for all decay modes.

\begin{table*}[htbp]
\centering
\caption{Summary of signal efficiency, expected background events and data candidates for each decay mode and each data taking period (SK-I to SK-IV). The error values correspond to the statistical uncertainty of the MC sample. “Lower” and “Upper” stand for $P_{\rm tot}<100$\,MeV/c and $100<P_{\rm tot}<250$\,MeV/c, respectively. The data events remained in the $p\rightarrow e^+\mu^+\mu^-$ and $p\rightarrow e^-\mu^+\mu^+$ modes are the same event.}
\begin{tabular}{lcccccccccccccc} \hline  \hline
  & \multicolumn{4}{c}{Efficiency (\%)} & \hspace{5mm} & \multicolumn{4}{c}{Background (events)} & \hspace{5mm} & \multicolumn{4}{c}{Candidate (events)}  \\
 Modes & I & II & III & IV & \hspace{5mm} & I & II & III & IV & \hspace{5mm}  & I & II & III & IV \\ \hline
 $p\rightarrow e^+e^+e^-$  \\
 (Lower) & 22.7 & 19.8 & 23.1 & 22.4 & \hspace{5mm} & $<$0.01 & $<$0.01 & $<$0.01 & $<$0.01 & \hspace{5mm}  & 0 & 0 & 0 & 0 \\  
 (Upper) & 43.9 & 40.4 & 44.3 & 41.1 & \hspace{5mm} & 0.19$\pm$0.04 & 0.10$\pm$0.02 & 0.05$\pm$0.01 & 0.24$\pm$0.07 & \hspace{5mm}  & 0 & 0 & 0 & 0 \\ 
 $p\rightarrow \mu^+e^+e^-$   \\
 (Lower) & 15.0 & 13.5 & 16.3 & 17.6 & \hspace{5mm} & 0.02$\pm$0.01 & 0.02$\pm$0.01 & 0.01$\pm$0.00 & $<$0.01 & \hspace{5mm}  & 0 & 0 & 0 & 0 \\
 (Upper) & 27.1 & 26.0 & 27.3 & 30.3 & \hspace{5mm} & 0.13$\pm$0.03 & 0.10$\pm$0.02 & 0.05$\pm$0.01 & 0.17$\pm$0.05 & \hspace{5mm}  & 0 & 0 & 0 & 0 \\
 $p\rightarrow \mu^-e^+e^+$   \\
 (Lower) & 11.9 & 11.1 & 12.6 & 14.9 & \hspace{5mm} & 0.02$\pm$0.01 & 0.02$\pm$0.01 & 0.01$\pm$0.00 & $<$0.01 & \hspace{5mm}  & 0 & 0 & 0 & 0 \\
  (Upper) & 20.8 & 19.8 & 22.3 & 25.9 & \hspace{5mm} & 0.13$\pm$0.03 & 0.10$\pm$0.02 & 0.05$\pm$0.01 & 0.17$\pm$0.05 & \hspace{5mm}  & 0 & 0 & 0 & 0 \\
  $p\rightarrow e^+\mu^+\mu^-$  \\
  (Lower) & 9.2 & 8.1 & 9.1 & 11.7 & \hspace{5mm} & $<$0.01 & $<$0.01 & $<$0.01 & $<$0.01 & \hspace{5mm}  & 0 & 0 & 0 & 0 \\
  (Upper) & 15.8 & 14.1 & 16.2 & 20.9 & \hspace{5mm} & 0.09$\pm$0.02 & 0.07$\pm$0.02 & 0.03$\pm$0.01 & 0.08$\pm$0.03 & \hspace{5mm}  & 0 & 0 & 0 & 1 \\
   $p\rightarrow e^-\mu^+\mu^+$  \\
  (Lower) & 11.1 & 10.9 & 11.9 & 14.4 & \hspace{5mm} & $<$0.01 & $<$0.01 & $<$0.01 & $<$0.01 & \hspace{5mm}  & 0 & 0 & 0 & 0  \\
  (Upper) & 19.9 & 18.2 & 20.0 & 24.2 & \hspace{5mm} & 0.09$\pm$0.02 & 0.07$\pm$0.02 & 0.03$\pm$0.01 & 0.08$\pm$0.03 & \hspace{5mm}  & 0 & 0 & 0 & 1  \\
  $p\rightarrow \mu^+\mu^+\mu^-$  \\
  (Lower) & 10.8 & 10.4 & 12.0 & 12.2 & \hspace{5mm} & $<$0.01 & $<$0.01 & $<$0.01 & $<$0.01 & \hspace{5mm}  & 0 & 0 & 0 & 0 \\
  (Upper) & 19.9 & 17.2 & 20.4 & 20.4 & \hspace{5mm} & 0.10$\pm$0.03 & 0.05 $\pm$0.01 & 0.03$\pm$0.01 & 0.22$\pm$0.06 & \hspace{5mm}  & 1 & 0 & 0 & 0  \\ \hline \hline
 \end{tabular}
\label{table_selection_summary}
\end{table*}

\begin{table*}[htbp]
\centering
\caption{The fraction [\%] of the background interaction modes remaining in the signal box for each proton decay channel. CC, NC, and QE stand for charged-current, neutral-current, and quasi-elastic neutrino interactions, respectively.}
\begin{tabular}{lcccc} \hline \hline
           & $p\rightarrow e^+e^+e^-$   & $p\rightarrow \mu^+e^+e^-$ & $p\rightarrow e^+\mu^+\mu^-$ & $p\rightarrow \mu^+\mu^+\mu^-$ \\
           &  & ($p\rightarrow \mu^-e^+e^+$)  & ($p\rightarrow e^-\mu^+\mu^+$) &  \\ \hline 
CCQE & 13  & 7  & 10  & 21 \\ 
CC single-$\pi$ & 41 & 32 & 34 & 58  \\
CC multi-$\pi$ & 12 & 27 & 42 & 13 \\
CC others & 13 & 12 & 5 & 4 \\
NC & 21 & 21 & 9 & 4 \\ \hline \hline
\end{tabular}
\label{table_bkg_fraction}
\end{table*}

\section{Systematic uncertainties}
Dominant systematic uncertainties for the signal efficiency are associated with the uncertainties in correlated decay and Fermi motion models.
Since the mechanism of correlated decay is not well understood, variation of the signal efficiency was evaluated with 0\% and 20\% probabilities of correlated decay compared to our nominal estimate of 10\%, and the spread was taken as the uncertainty. 
In the simulation of the signal, Fermi motion is simulated based on the electron-$^{12}$C scattering experiment \cite{exp_fermi_motion}.
On the other hand, the Fermi motion model for the background sample is based on the Fermi gas model.
This model difference is considered as a source of systematic uncertainty.

For the background, systematic uncertainties on the neutrino flux and cross section models are taken into account in the estimated background rates.
These uncertainties are estimated by an event-by-event weighting method based on the neutrino oscillation analysis in SK \cite{latest_oscillation}.
A pion generated by a neutrino interaction can interact with a nucleon in oxygen (final state interaction, FSI). 
It is also possible to interact with other nuclei in water after escaping the original nucleus (secondary interaction, SI).
FSI/SI are simulated by a pion cascade model and their uncertainties are taken into account.

Systematic uncertainties for the detector performance and reconstruction are taken into account for both signal and background.
In order to estimate these uncertainties, control sample data and MC are
compared for each source of systematic uncertainties.
We consider uncertainties for FV, detector non-uniformity, energy scale, ring counting, PID, decay electron tagging and neutron tagging.
Systematic uncertainty for the detector exposure is negligible.
We assigned a 1\% error for the detector exposure to be conservative. 

Systematic uncertainties for the signal and background are summarized in TABLE \ref{table_syst_sig} and TABLE \ref{table_syst_bkg}, respectively.
The dominant uncertainties for the background due to the event reconstruction are energy scale and detector non-uniformity.
Uncertainties of the energy scale \cite{latest_oscillation} are taken into account for all the charge-related reconstruction parameters. 
The effect of detector non-uniformity of the energy scale \cite{latest_oscillation} is taken into account for the total momentum reconstruction.

The dominant error for $p\rightarrow\mu^+\mu^+\mu^-$ (TABLE \ref{table_syst_sig}) comes from uncertainty of the decay electron tagging. 
Since the number of candidate events with 3 $\mu$-like rings and 2 or 3 decay electrons (selections for the $p\rightarrow\mu^+\mu^+\mu^-$) is smaller than for the other modes, the statistical error of the atmospheric neutrino control sample data used to estimate the systematic error is larger.

\begin{table*}[tbp]
\centering
 \caption{Summary of systematic uncertainty [\%] for the signal averaged over the live time of each period. “Lower” and “Upper” stand for $P_{\rm tot}<100$\,MeV/c and $100<P_{\rm tot}<250$\,MeV/c, respectively. }
\begin{tabular}{lcccc} \hline \hline
 Modes & Correlated decay & Fermi momentum & Detector \& Reconstruction & Total \\ \hline
  $p\rightarrow e^+e^+e^-$  \\
  (Lower) & 4.0 & 10.4 & 5.9 & 12.6 \\
  (Upper) & 9.3 & 3.1 & 4.4 & 10.8 \\
   $p\rightarrow \mu^+e^+e^-$  \\
   (Lower) & 3.9 & 10.3 & 8.1 & 13.7 \\
   (Upper) & 9.4 & 3.4 & 7.5 & 12.6 \\
   $p\rightarrow \mu^-e^+e^+$  \\
   (Lower) & 3.9 & 10.3 & 8.1 & 13.7 \\
   (Upper) & 9.6 & 3.0 & 7.6 & 12.5 \\
   $p\rightarrow e^+\mu^+\mu^-$  \\
   (Lower) & 3.7 & 10.1 & 8.3 & 13.6 \\
   (Upper) & 9.5 & 3.5 & 7.2 & 12.7 \\
   $p\rightarrow e^-\mu^+\mu^+$  \\
   (Lower) & 3.7 & 9.4 & 8.0 & 13.1 \\
   (Upper) & 8.8 & 5.6 & 7.2 & 12.9 \\
   $p\rightarrow \mu^+\mu^+\mu^-$  \\
   (Lower) & 3.8 & 10.5 & 18.9 & 22.1 \\
   (Upper) & 9.7 & 6.5 & 18.6 & 22.1 \\ \hline \hline
 \end{tabular}
\label{table_syst_sig}
\end{table*}

\begin{table*}[tbp]
\centering
\caption{Summary of systematic uncertainty [\%] for the background averaged over the live time of each period. }
\begin{tabular}{lccccc} \hline \hline
 & Neutrino  & Neutrino & Pion & Detector \& &  \\ 
 Modes & flux & cross section & FSI/SI & Reconstruction & Total \\ \hline
  $p\rightarrow e^+e^+e^-$  & 7.0 & 14.1 & 1.9 & 32.4 & 36.1 \\
  $p\rightarrow \mu^+e^+e^-$ ($p\rightarrow \mu^-e^+e^+$) & 7.3 & 16.9 & 1.9 & 19.3 & 26.7 \\
  $p\rightarrow e^+\mu^+\mu^-$ ($p\rightarrow e^-\mu^+\mu^+$) & 8.2 & 23.6 & 3.3 & 19.6 & 32.0 \\
  $p\rightarrow \mu^+\mu^+\mu^-$ & 8.3 & 16.6 & 1.8 & 32.4 & 37.4 \\ \hline \hline
  \end{tabular}
\label{table_syst_bkg}
\end{table*}

\section{Result}
\begin{figure*}[htbp]
	\centering
	\includegraphics[width=140mm]{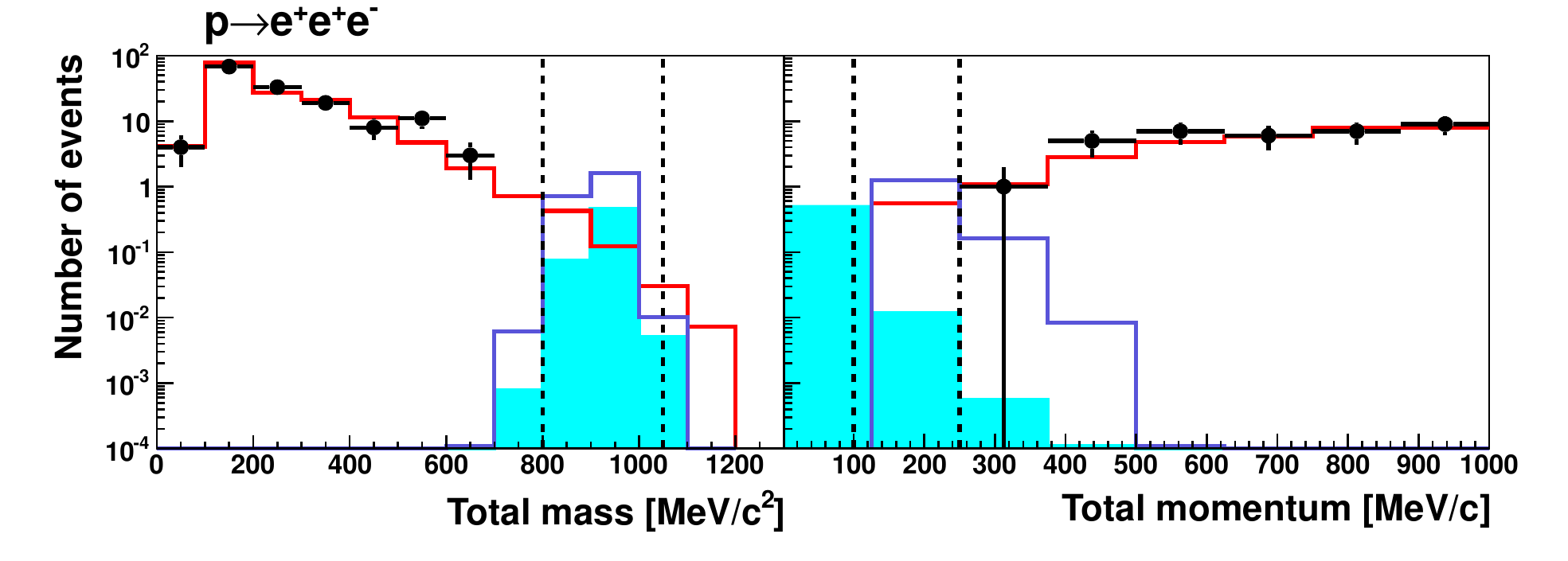}
	\includegraphics[width=140mm]{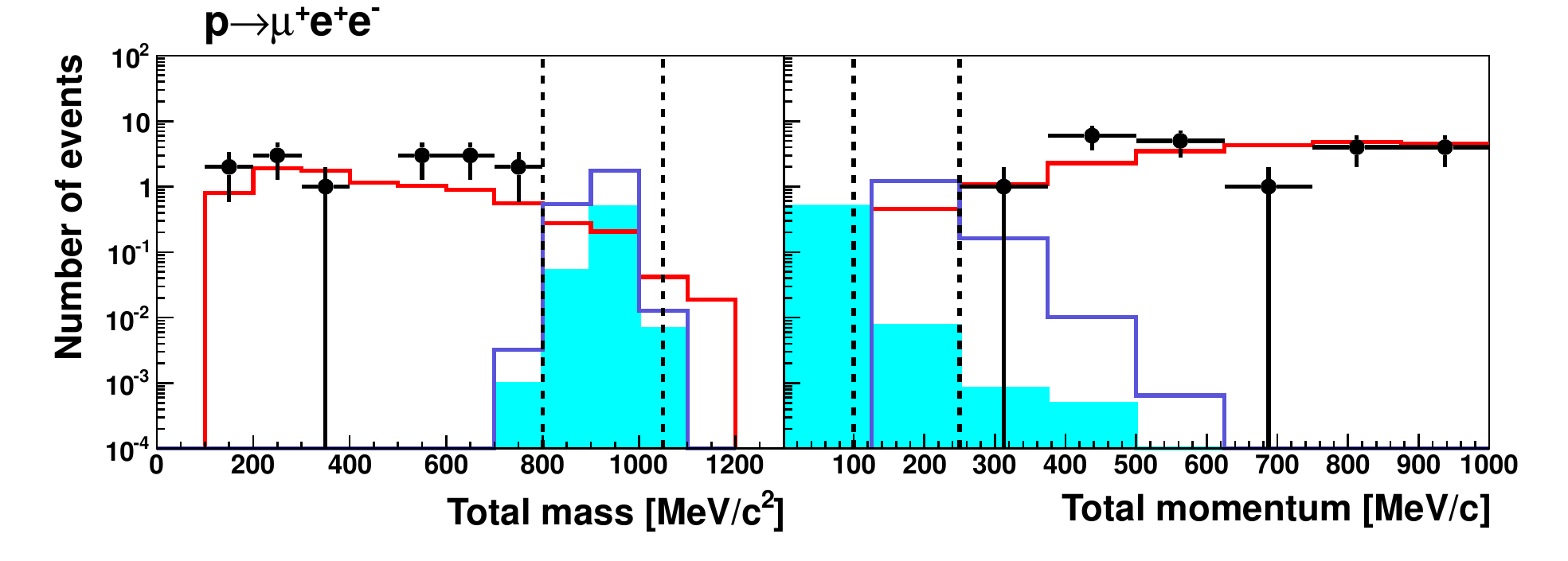}
	\includegraphics[width=140mm]{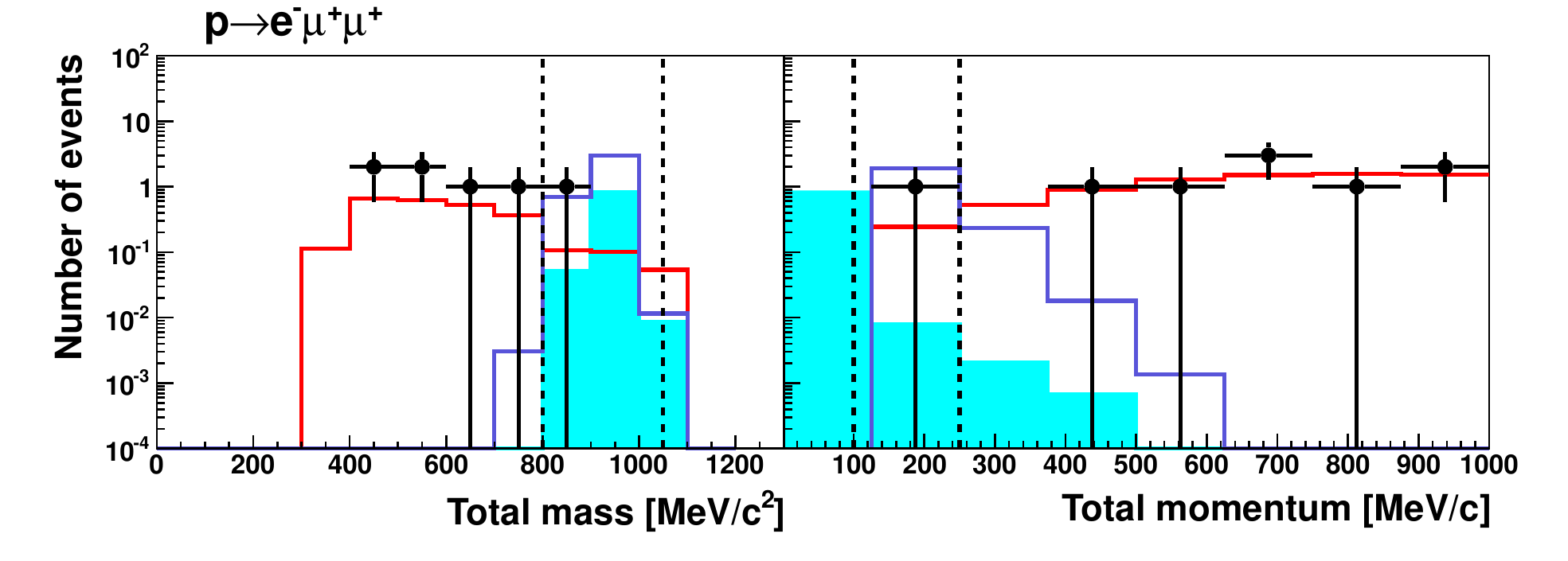}
	\includegraphics[width=140mm]{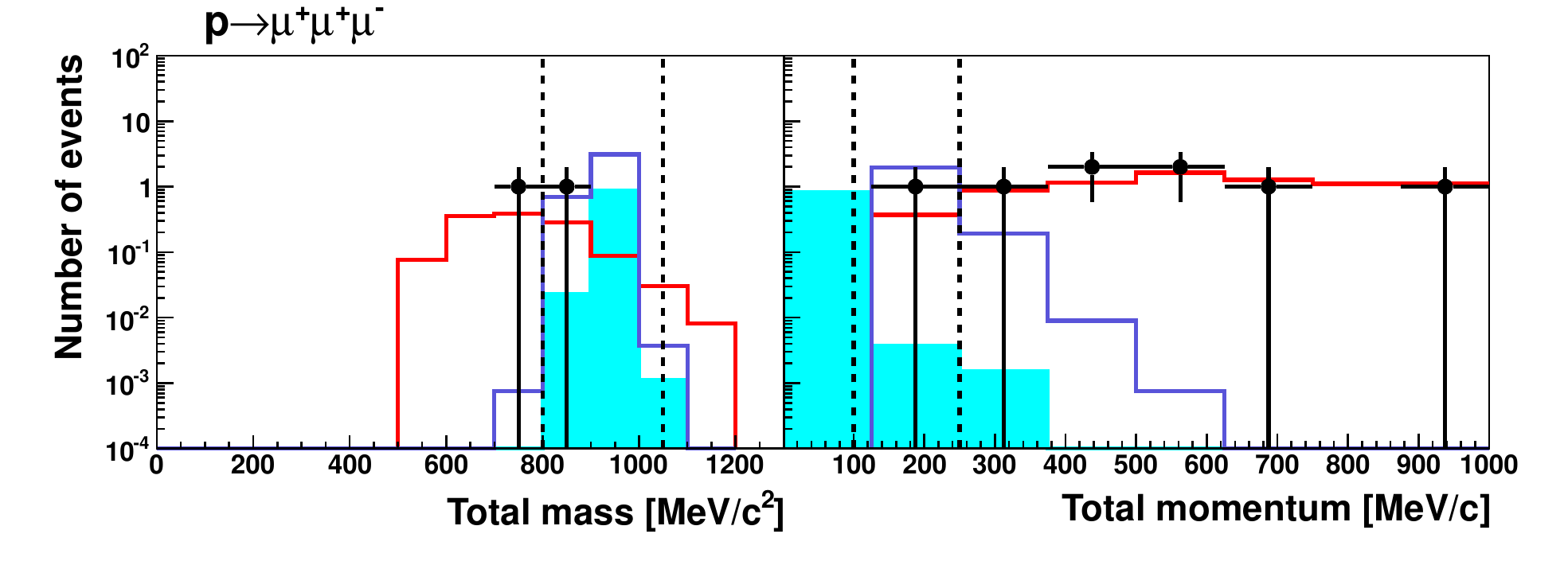}
	\caption{Data (black dots) and background (red line) comparison for total mass (left) and momentum (right)  after the selections C1-C3 and C5 are applied. The expected signal (free proton only) distribution is shown by the blue line (filled cyan histogram). Additionally a $P_{\rm tot}<250$\,MeV/c cut is applied on the total mass plot and a $800<M_{\rm tot}<1050$\,MeV/c$^2$ cut is applied on the total momentum plot. From the top to the bottom, the $p\rightarrow e^+e^+e^-$, $p\rightarrow \mu^+e^+e^-$, $p\rightarrow e^-\mu^+\mu^+$ and $p\rightarrow \mu^+\mu^+\mu^-$ modes are shown, respectively. Dotted black lines show the boundary of the signal box. SK-I-IV are combined in signal, background MC and data. Background MC is normalized by atmospheric neutrino flux, oscillation probability and livetime. Signal MC is normalized to the partial proton lifetime limit calculated in section \ref{section_lifetime}.
The number of signal MC events for the  $p\rightarrow\mu^-e^+e^+$ mode is lower than that of  $p\rightarrow\mu^+e^+e^-$ by 19\%, due to the different effective lifetimes (and therefore different decay electron production probabilities) of the $\mu^{-}$ and $\mu^{+}$ in water.
Similarly, the signal MC for $p\rightarrow e^+\mu^+\mu^-$ has 20\% fewer events than that for the $p\rightarrow e^-\mu^+\mu^+$ mode. 
As in FIG.\ref{fig_cutflow}, the data and background MC figures are the same for modes that differ only by the charge sign of the leptons.
}
	\label{fig_1D}
\end{figure*}

No events are found in the signal box region for the $p\rightarrow e^+e^+e^-$ and $p\rightarrow \mu^+e^+e^-$ ($p\rightarrow \mu^-e^+e^+$) modes.
One event is observed in the upper signal box for both the $p\rightarrow e^+\mu^+\mu^-$ ($p\rightarrow e^-\mu^+\mu^+$) and $p\rightarrow \mu^+\mu^+\mu^-$modes (all the event displays in \cite{event_displays}).
Observed candidates are summarized in TABLE \ref{table_selection_summary} and TABLE \ref{table_limit}.
FIG.\ref{fig_cutflow} shows the comparisons of the number of observed events to the estimated background along the event selection for each proton decay mode.
The observed numbers of events at each cut step are consistent with the expected background. 
Total mass and momentum distributions are shown in FIG. \ref{fig_2D} as two dimensional plots and in FIG. \ref{fig_1D} as histograms.
Data and background MC distributions agree well in each decay channel.

The candidate for the $p\rightarrow e^+\mu^+\mu^-$ ($p\rightarrow e^-\mu^+\mu^+$) modes is observed in the upper signal box of the SK-IV period.
Reconstructed proton mass and momentum for this candidate are 882\,MeV/c$^2$ and 160\,MeV/c, respectively.
Another candidate for the $p\rightarrow \mu^+\mu^+\mu^-$ mode is found in the upper signal box of the SK-I period.
There are two decay electrons, and the total mass and momentum are 835\,MeV/c$^2$ and 170\,MeV/c, respectively.
These events were visually inspected and they appear not to be mis-reconstruction events.

The expected background events in the $p\rightarrow e^+\mu^+\mu^-$ ($p\rightarrow e^-\mu^+\mu^+$) mode is 0.27$\pm$0.04 (stat.) events.
Assuming a Poisson distribution of mean 0.27, the probability to observe $\geq1$ events is 18.4\%.
Considering the expected background events in the $p\rightarrow \mu^+\mu^+\mu^-$ to be 0.40$\pm$0.07 (stat.) events, the probability of observing $\geq1$ events is 25.8\%.


\begin{table}[htbp]
\centering
 \caption{Summary of expected background events with statistical errors, number of candidates, Poisson probabilities to observe events greater than or equal to the number of data candidates, and estimated partial lifetime lower limits.}
\begin{tabular}{lcccc} \hline \hline
 & & & & Lifetime limit \\
 & Background  & Candidate & Probability & ($\times10^{34}$ years)    \\ 
 Modes & (events) & (events) & (\%) & at 90\% CL   \\ \hline
$p\rightarrow e^+e^+e^-$  & 0.58$\pm$0.08 & 0 & - & 3.4  \\
$p\rightarrow \mu^+e^+e^-$  & 0.50$\pm$0.06 & 0 & - & 2.3 \\
$p\rightarrow \mu^-e^+e^+$  & 0.50$\pm$0.06 & 0 & - & 1.9 \\
$p\rightarrow e^+\mu^+\mu^-$  & 0.27$\pm$0.04 & 1 & 18.4 & 0.92 \\
$p\rightarrow e^-\mu^+\mu^+$  & 0.27$\pm$0.04 & 1 & 18.4 & 1.1 \\
$p\rightarrow \mu^+\mu^+\mu^-$  & 0.40$\pm$0.07 & 1 & 25.8 & 1.0 \\ \hline \hline
  \end{tabular}
\label{table_limit}
\end{table}

\section{Partial Lifetime limit \label{section_lifetime}}
The observed events are consistent with expected backgrounds; therefore lower proton lifetime limits at 90\% confidence level (CL) with respect to each proton decay mode are calculated by using a Bayesian method \cite{bayesian,bayesian2}.
The limit calculation is the same as for recent nucleon decay analyses \cite{pdk_sk_epi, pdk_sk_nuk,pdk_sk_antilepton_meson}.
We have 8 signal regions (4 periods $\times$ 2 boxes) for each decay mode.
The probability density function (PDF) is defined for each region as below.
\begin{eqnarray}
P(\Gamma|n_i) = \frac{1}{A_i} \iiint \frac{e^{-(\Gamma \lambda_i \epsilon_i + b_i)} (\Gamma \lambda_i \epsilon_i + b_i)^{n_i}}{n_i!} \times \\ \nonumber
 P(\Gamma)P(\lambda_i)P(\epsilon_i)P(b_i) d\epsilon_i d\lambda_i db_i
\end{eqnarray}
Here, $i$ is the index of each signal region, $A_i$ is a normalization factor, $\Gamma$ is the decay rate, $n_i$ is the observed events, $\lambda_i$ is the exposure, $\epsilon_i$ is the signal efficiency and $b_i$ is the expected background events.
$P(\Gamma)$ represents the probability distribution for the decay rate, assumed to be uniform.
$P(\lambda_i)$ and $P(\epsilon_i)$ are the probabilities for the exposure and signal efficiency, respectively, described by a Gaussian. 
$P(b_i)$ is the probability for the expected background defined by the convolution of Gaussian and Poisson distributions.
All PDFs are combined and the upper limits of the decay rate $\Gamma_{\rm limit}$ at 90\% CL are estimated as follows.
\begin{equation}
\int_{\Gamma=0}^{\Gamma_{\rm limit}} \prod_{i=1}^{8} P(\Gamma|n_i) d\Gamma = 0.9
\end{equation}
Finally the lower limit of partial proton lifetime is calculated according to
\begin{equation}
\tau/B = \frac{1}{\Gamma_{\rm limit}},
\end{equation}
Here $B$ is the branching ratio of each proton decay mode.
By using these functions, the partial lifetime limits at 90\% CL for each mode of proton decay into three charged leptons are calculated as summarized in TABLE \ref{table_limit}.

\section{Conclusion}
\begin{figure}[htbp]
\centering
\includegraphics[width=80mm]{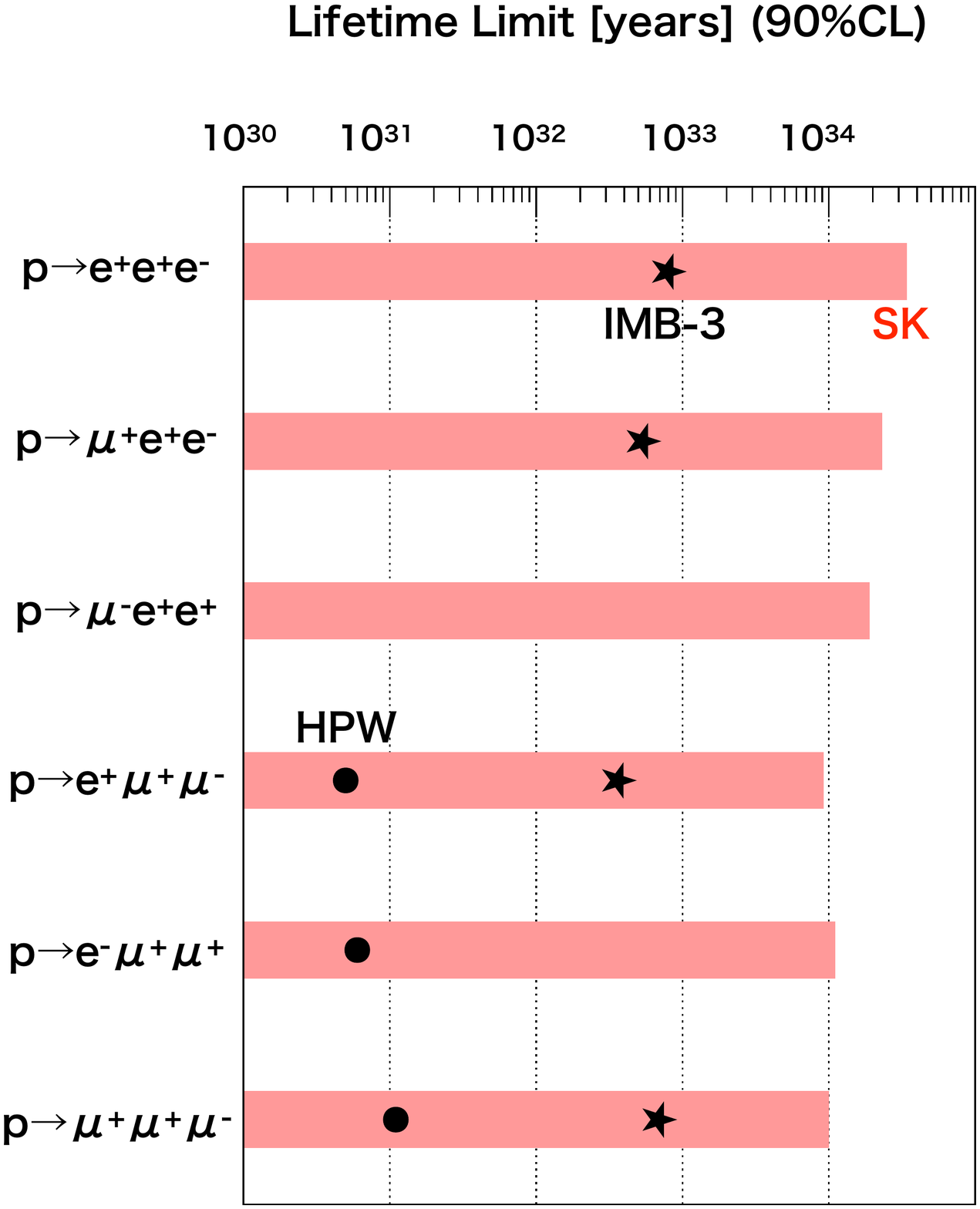}
\caption{Estimated partial lifetime lower limit for each mode of proton decay into three charged leptons by this analysis (red histogram), IMB-3 (star)\cite{IMB_proton_decay} and HPW (circle)\cite{HPW_proton_decay} experiments. The $p\rightarrow\mu^-e^+e^+$ channel was not searched for by IMB-3 and HPW experiments.}
\label{fig_compare_limit}
\end{figure}

Proton decay into three charged leptons was searched for by using 0.37\,Mton$\cdot$years of data collected by SK.
The observed data were consistent with the atmospheric neutrino background prediction and no clear indication of proton decay was observed.
According to the observation, the model~\cite{p_charged_leptons} for these decay modes at an energy scale below 100\,TeV was excluded by this analysis.
The lower partial lifetime limits at 90\% CL were calculated for each mode as summarized in TABLE \ref{table_limit}.
Compared with the previous limits by IMB-3 and HPW experiments, each limit was improved by 15-1800 times in this analysis as shown in FIG.\ref{fig_compare_limit}.
A first limit has been set for the $p\rightarrow\mu^-e^+e^+$ mode.

\section*{Acknowledgements}


We gratefully acknowledge the cooperation of the Kamioka Mining and Smelting Company. The Super-Kamiokande experiment has been built and operated from funding by the Japanese Ministry of Education, Culture, Sports, Science and Technology, the U.S. Department of Energy, and the U.S. National Science Foundation. Some of us have been supported by funds from the National Research Foundation of Korea NRF-2009-0083526 (KNRC) funded by the Ministry of Science, ICT, and Future Planning and the the Ministry of Education (2018R1D1A3B07050696, 2018R1D1A1B07049158), the Japan Society for the Promotion of Science, the National Natural Science Foundation of China under Grants No. 11235006, the Spanish Ministry of Science, Universities and Innovation (grant PGC2018-099388-B-I00), the National Science and Engineering Research Council (NSERC) of Canada, the Scinet and Westgrid consortia of Compute Canada, the National Science Centre, Poland (2015/18/E/ST2/00758), the Science and Technology Facilities Council (STFC) and GridPPP, UK, the European Union's H2020-MSCA-RISE-2018 JENNIFER2 grant agreement no.822070, and  H2020-MSCA-RISE-2019 SK2HK grant agreement no. 872549.

\end{document}